\documentclass[aps,prd,longbibliography,nofootinbib,10pt]{revtex4-2}

\usepackage[T1]{fontenc}
\usepackage{microtype}
\usepackage{graphicx}
\graphicspath{{figures/}}
\usepackage{dcolumn}
\usepackage{amsmath,amssymb,mathtools}
\usepackage{lmodern}
\usepackage{booktabs}
\usepackage{array}
\usepackage{tabularx}
\newcolumntype{Y}{>{\raggedright\arraybackslash}X}
\newcolumntype{P}[1]{>{\raggedright\arraybackslash}p{#1}}
\newcolumntype{d}[1]{D{.}{.}{#1}}
\usepackage{hyperref}
\hypersetup{hidelinks}
\hypersetup{colorlinks=true,linkcolor=blue,citecolor=blue,urlcolor=blue}
\newcommand{\NdotC}{\dot N_{\rm C}}
\newcommand{\ndotC}{\dot n_{\rm C}}
\newcommand{\Pstar}{P_{\star}^{\rm inc}}
\newcommand{\Pint}{P_{\rm int}}
\newcommand{\Pabs}{P_{\rm abs}}
\newcommand{\PX}{P_X}
\newcommand{\Pbio}{P_{\rm bio}}
\newcommand{\Pmaint}{P_{\rm maint}}
\newcommand{\Pinf}{P_{\rm info}}
\newcommand{\Pprod}{P_{\rm prod}}
\newcommand{\phibio}{\phi_{\rm bio}}
\newcommand{\phimaint}{\phi_{\rm maint}}
\newcommand{\jcopy}{j_{\rm copy}}
\newcommand{\Ncopy}{\dot N_{\rm copy}}
\newcommand{\kb}{k_{\rm B}}
\newcommand{\Tenv}{T}
\newcommand{\DGassim}{\Delta G_{\rm assim}}
\newcommand{\etabio}{\eta_{\rm bio}}
\newcommand{\Isite}{I_{\rm site}}

\newcommand{\esite}{e_{\rm site}}
\newcommand{\muerr}{\mu}
\newcommand{\kp}{k_p}
\newcommand{\pos}[1]{\left[#1\right]_{+}}

\begin{document}

\title{An Information--Thermodynamic Accounting Framework for \\ Planetary Biosphere Productivity}

\author{Slava G. Turyshev}
\affiliation{Jet Propulsion Laboratory, California Institute of Technology,\\
4800 Oak Grove Drive, Pasadena, CA 91109-0899, USA}
\date{October 23, 2026}

\begin{abstract}
Planetary primary producers use stellar or internally generated free energy to maintain biomass and produce new material. We formulate the associated productivity ceiling while separating environmental losses, basal maintenance, retained biomass energy and net dissipation from heritable copying. Maintenance copy-reset cycles and copies retained in new biomass have distinct thermodynamic costs. For a specified feasible producer population, the normalized ceiling is $(1-x_0)/(1+q)$, where $x_0$ is the baseline copying fraction and $q$ is the growth-associated copying cost relative to other production costs. Exact interval bounds and global sensitivity analyses quantify both effects without assigning empirical probabilities to unmeasured biological parameters. Observation-constrained planetary resources and strain-matched genome and cellular-carbon measurements provide separate physical and biological benchmarks. For the two mean inventories, a 10\% growth-copying correction requires an effective copying cost $e_g\chi$ about 63--84 times a published DNA fuel benchmark at unit production efficiency; the benchmark is not a measured dissipation cost. We also show that partitioning an already measured total maintenance budget cannot change the absolute productivity ceiling, and that population adjustment can preserve information-load fractions as supply falls. The framework provides conditional feasibility tests, inverse thresholds and measurement priorities, rather than a calibrated prediction of extraterrestrial productivity.
\end{abstract}

\maketitle

%***********************************************
\section{Introduction}
\label{sec:intro}

Planets receive energy from their host stars and from internal sources, including radioactive decay, residual heat and tidal dissipation. Biological primary producers use part of the available energy to build organic matter from inorganic substrates. On Earth, photosynthetic organisms capture light, while chemical primary producers exploit reactions between substances maintained away from equilibrium. Energy is also needed to sustain existing cells, replace damaged components and copy the templates required for inheritance. The resulting production of new biomass, after the producers' own respiratory losses, is net primary productivity (NPP); it is a material production rate, not necessarily an increase in the standing biomass of an ecosystem \cite{Field1998,Chapin2006}.

The amount of energy entering a planet does not by itself determine this productivity. Useful work requires an appropriate energy contrast: for example, radiation absorbed by a photochemical system or a favorable chemical reaction supplied with reactants. We use \emph{exergy} for the maximum work available relative to a specified local environment \cite{Kleidon2010,KrissansenTotton2016}. Only some of that resource reaches biological producers. This distinction supplements climate-based habitability criteria \cite{Kasting1993,Kopparapu2013} and is essential for subsurface oceans and free-floating planets, where internal heat need not provide an accessible chemical energy supply.

Copying and maintaining heritable templates add a further allocation question: how much of the accessible free energy can they consume before biomass production is materially reduced? Information thermodynamics constrains specified copying and resetting processes \cite{Landauer1961,Bennett1982,Parrondo2015,Ouldridge2017PRX,Ouldridge2017PRL}. It does not supply a universal dissipative cost per copied nucleotide. Cellular fuel budgets and astrobiological information measures address related questions \cite{Kempes2017,Lynch2015,Lorenz2004,Scharf2024}, but their planetary application requires a consistent boundary separating energy retained in new biomass from energy dissipated during its production.

This paper develops the power--productivity part of the constants-to-biospheres framework in Ref.~\cite{Turyshev2026Pathway}, whose Eq.~(117) already gives a copying-corrected productivity ceiling. The present contributions are the primary-producer accounting boundary, separate maintenance and growth-copying costs, and explicit closures relating copying to standing biomass and new production. We derive exact reduction factors, feasibility conditions and uncertainty bounds, and distinguish measurable total demand from information-specific cost attribution. These results identify which conclusions survive uncertain biological inputs and which require additional measurements.

Planetary observations constrain energy sources more directly than biological capture, repair or copying dissipation \cite{Schwieterman2018,Meadows2017,Olson2018}. Accordingly, measured quantities, model-derived constraints and scenario choices are kept separate. We use strain-matched measurements for a terrestrial reference producer, propagate their stated uncertainties, and examine the remaining unknowns through threshold calculations and declared sensitivity domains. No probability distribution for alien physiology is inferred from those domains. Table~\ref{tab:symbols} introduces the principal quantities; Secs.~\ref{sec:power}--\ref{sec:bound} develop the accounting, Secs.~\ref{sec:examples}--\ref{sec:retrieval} apply the tests, and the appendices document the parameter evidence, global sensitivity and population closure.

\begin{table}[t]
\caption{Principal quantities in the accounting. All powers use the same environmental free-energy reference. Cellular variables are defined with their closures in Sec.~\ref{sec:bound}.}
\label{tab:symbols}
\centering
\begin{tabular}{lll}
\toprule
Quantity & Meaning & Units \\
\midrule
$\Pstar,\Pint$ & Intercepted stellar power; internal energy supply & W \\
$\Pabs,\PX,\Pbio$ & Absorbed power; exergy; supply delivered to biological interfaces & W \\
$\Pmaint=P_{\rm env}+P_0$ & Environmental overhead and non-information basal maintenance & W \\
$\Pinf,\Pprod$ & Net copying dissipation; productive metabolic power & W \\
$\NdotC$ & Net primary productivity of the specified producers & mol C s$^{-1}$ \\
$\DGassim$ & Free energy retained per unit biomass carbon & J mol C$^{-1}$ \\
$\etabio$ & Production efficiency, excluding separately charged costs & dimensionless \\
$e_0,e_g$ & Net dissipation per maintenance-cycle or growth-copying site & J site$^{-1}$ \\
$P_{\rm res}$ & Residual supply, $\Pbio-\Pmaint$ & W \\
$\Theta_{\rm info}$ & Copying fraction, $\Pinf/P_{\rm res}$ & dimensionless \\
\bottomrule
\end{tabular}
\end{table}

%***********************************************
\section{Power reservoirs, exergy and material NPP}
\label{sec:power}

We distinguish the physical power reservoirs from the biological allocation of their usable free energy.  The stellar power intercepted by a planet is
\begin{equation}
  \Pstar = \pi R_p^2 S_\star,
  \label{eq:pstar}
\end{equation}
where $R_p$ is planetary radius and $S_\star$ is bolometric flux at the orbit.  Internal sources such as radiogenic heating, primordial cooling and tidal dissipation contribute $\Pint$.  The physically absorbed power is
\begin{equation}
  \Pabs = (1-A_B)\Pstar + \Pint,
  \label{eq:pabs}
\end{equation}
with $A_B$ the Bond albedo.  Not all absorbed power can do useful work.  Let
\begin{equation}
  \PX = f_X(\Tenv,\{\mu_i\},{\cal G},{\cal S})\,\Pabs
  \label{eq:exergy}
\end{equation}
be the rate at which the planet generates exergy relative to the relevant local environment.  The factor $f_X$ is not universal.  It depends on source and sink temperatures, chemical potentials $\{\mu_i\}$, geometry ${\cal G}$, spectral distribution and the configuration of reservoirs ${\cal S}$.  For heat engines, $f_X$ is bounded by a Carnot factor.  For photochemistry or redox chemistry it is set by photon spectra, reaction affinities, transport and reactant supply.  Thus heat flux by itself is not $\Pbio$. Equations~(\ref{eq:pabs})--(\ref{eq:exergy}) describe a sustained, time-averaged resource budget. Depletion of a previously stored chemical reservoir or an additional external energy source must enter as an explicit extra supply; it cannot be attributed to the contemporaneous heat flux.

Only a subset of $\PX$ may reach the habitable interface in a form accessible to life.  We define
\begin{equation}
  \Pbio = f_{\rm bio}\PX,
  \qquad
  \phibio = \frac{\Pbio}{A_{\rm hab}},
  \label{eq:pbio}
\end{equation}
where $0\le f_{\rm bio}\le 1$ and $A_{\rm hab}$ is the area over which the fluxes are averaged.  Here $\Pbio$ is the free-energy supply delivered to biologically accessible interfaces before the explicitly identified downstream losses.  Transport or chemical losses included in $f_{\rm bio}$ must not also be charged as overheads.

The biological control volume comprises the primary producers whose biomass output defines NPP.  Their exported biomass supplies consumers and decomposers downstream; consumer copying is not an additional parallel subtraction from the same input budget.  NPP is a production flux and need not equal net ecosystem biomass accumulation \cite{Field1998,Chapin2006}.  All work, dissipation and retained biomass free energy are evaluated relative to the same environmental reference.

The producer budget, including unused supply $P_u\ge0$, is
\begin{equation}
  \Pbio = \Pmaint + \Pinf + \Pprod + P_u,
  \qquad \Pmaint=P_{\rm env}+P_0.
  \label{eq:partition}
\end{equation}
Here $P_{\rm env}$ denotes specified losses or work needed to sustain the assumed habitable interface after the delivery represented by Eq.~(\ref{eq:pbio}).  $P_0$ is non-information basal maintenance of the standing producer population, including membrane homeostasis and ordinary protein renewal.  Neither term includes raw heat lacking usable exergy.  Growth-associated non-information losses enter $\etabio$ below.  This separation permits a nonzero maintenance demand at negligible growth; it cannot generally be represented by a constant growth efficiency alone.  Each process is assigned to exactly one term.

The residual productive budget is
\begin{equation}
  P_{\rm res}\equiv\Pbio-\Pmaint=(1-f_{\rm maint})\Pbio,
  \qquad f_{\rm maint}\equiv\Pmaint/\Pbio .
  \label{eq:pres}
\end{equation}
The fraction $f_{\rm maint}$ is a budget ratio, not a universal physiological parameter. Measurements of total cellular maintenance generally include repair; such a total cannot be assigned to $P_0$ and then charged for repair again. Section~\ref{sec:identifiability} gives a consistent treatment when only total maintenance is constrained.

The productive term supplies the net output of new primary-producer biomass and the growth-associated losses represented by $0<\etabio\le1$:
\begin{equation}
  \NdotC = \frac{\etabio\Pprod}{\DGassim},
  \qquad
  \Pprod = \frac{\NdotC\DGassim}{\etabio}.
  \label{eq:npp_def}
\end{equation}
The symbol $\NdotC$ has units mol C s$^{-1}$; its areal version is $\ndotC=\NdotC/A_{\rm hab}$.  The positive quantity $\DGassim$ is the retained free energy per mole of biomass carbon relative to the specified chemical reference, including retained structural and informational contributions.  Its value is composition- and environment-dependent.  Thus $\NdotC\DGassim$ is a biomass free-energy production flux.  The difference $\Pprod-\NdotC\DGassim$ accounts only for non-information production losses: $\etabio$ excludes basal and copying costs already charged in Eq.~(\ref{eq:partition}).  These definitions prevent the same expenditure from entering both an efficiency and an explicit power term.

\begin{figure}[t]
\centering
\includegraphics[width=0.95\linewidth]{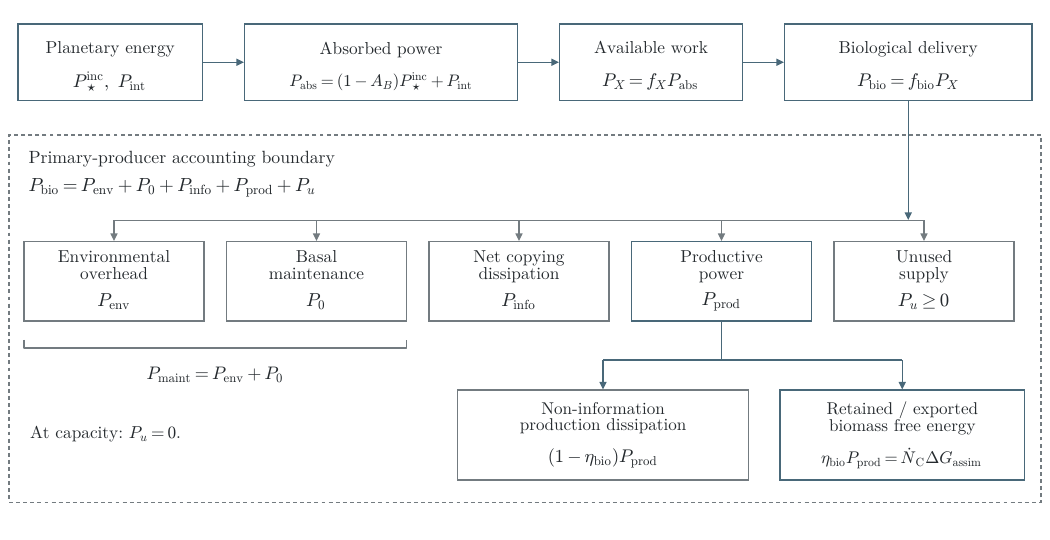}
\caption{Power accounting at the primary-producer boundary. Physical sources determine absorption, available work and biological delivery. The general budget includes unused supply $P_u\ge0$; the productivity ceiling uses $P_u=0$. Environmental and non-information basal demands form $\Pmaint=P_{\rm env}+P_0$. Productive power subdivides into non-information production dissipation and retained or exported biomass free energy. These are components of $\Pprod$, not additional deductions. Exported biomass can support downstream consumers. Arrows indicate accounting relationships and are not scaled to power.}
\label{fig:architecture}
\end{figure}

Figure~\ref{fig:architecture} summarizes the accounting boundary.

%***********************************************
\section{Information thermodynamics of copying and proofreading}
\label{sec:info_thermo}

Persistent templated inheritance requires both maintenance of existing templates and copying associated with new biomass.  These processes have different thermodynamic endpoints.  A maintenance copy-use-reset cycle returns its material inventory to the reference state; a growth-associated copy can leave the producer compartment as retained biomass.  Correlation free energy retained in that output must not also be counted as dissipated copying power \cite{Ouldridge2017PRX,Ouldridge2017PRL,Poulton2019}.

For a memoryless, symmetric $\kp$-letter channel with per-site error probability $\muerr$, the conditional entropy is
\begin{equation}
  H_{\kp}(\muerr)=h_2(\muerr)+\muerr\log_2(\kp-1),
  \label{eq:Hk}
\end{equation}
where $h_2(\muerr)=-\muerr\log_2\muerr-(1-\muerr)\log_2(1-\muerr)$.  For equiprobable template symbols the output is also uniform, giving
\begin{equation}
  \Isite=\log_2\kp-H_{\kp}(\muerr).
  \label{eq:isite}
\end{equation}
For biased symbols one instead uses $I(X;Y)=H(Y)-H(Y|X)$; correlations between successive sites require the corresponding sequence information rate.  High-fidelity, uniform DNA-like copying has $\kp=4$ and $\Isite\simeq2$ bits per nucleotide site.

At fixed marginal reference states, template-copy correlations carry free energy $\kb T\ln2\,\Isite$, where $\kb$ is Boltzmann's constant and $T$ the local bath temperature.  If a maintenance cycle destroys those correlations without recovering useful work, the associated lost-work contribution is
\begin{equation}
  e_{\rm corr}=\kb T\ln2\,\Isite.
  \label{eq:landauer_corr}
\end{equation}
The cycle, its resetting reservoirs and their restoration are included in this accounting.  Correlation creation alone is not an irreversible heat cost: reversible copying can retain the supplied work in its products.  Equation~(\ref{eq:landauer_corr}) is therefore applied only to the specified correlation-losing maintenance cycles, with material preparation and chemical free-energy changes accounted separately \cite{Ouldridge2017PRX,Ouldridge2017PRL}.

Proofreading costs also depend on the mechanism and its event fluxes.  For a chemically driven proofreading pathway,
\begin{equation}
 P_{\rm proof}^{\rm fuel}=J_{\rm proof}\Delta\mu_{\rm proof},
 \qquad w_{\rm proof}^{\rm fuel}=\frac{P_{\rm proof}^{\rm fuel}}{\dot N_{\rm out}},
 \label{eq:proof}
\end{equation}
where $J_{\rm proof}$ is the net proofreading-event rate, $\Delta\mu_{\rm proof}$ the supplied free energy per event, and $\dot N_{\rm out}$ the successful site-output rate.  Multiple driven pathways contribute a sum of such powers.  Accuracy bounds constrain chemical driving and product free-energy changes; the average cost per output site additionally depends on $J_{\rm proof}/\dot N_{\rm out}$ \cite{Sartori2015,Rao2015,Song2020}.  Fuel expenditure becomes a net dissipation cost only after retained free energy and any recovered work are removed.

Let $e_0$ denote net dissipation per completed maintenance-cycle site.  For the cycle specified above,
\begin{equation}
 e_0=e_{\rm corr}+e_{\rm irr},\qquad e_{\rm irr}\ge0,
 \label{eq:esite}
\end{equation}
where $e_{\rm irr}$ contains additional irreversibility without counting the same chemical expenditure twice.  Let $e_g\ge0$ instead denote excess dissipation per growth-associated copied site, excluding all free energy retained in new biomass and already included in $\DGassim$.  The maintenance-cycle floor does not impose an identical floor on $e_g$.  At $T=300$ K and $\Isite\simeq2$, $e_{\rm corr}=5.74\times10^{-21}$ J site$^{-1}$.  A biochemical fuel expenditure is not automatically equal to either $e_0$ or $e_g$; the distinction is retained in the quantitative comparisons below.

Writing the maintenance and growth site rates as $\dot N_{{\rm copy},0}$ and $\dot N_{{\rm copy},g}$ gives
\begin{equation}
 \Pinf=\dot N_{{\rm copy},0}e_0+\dot N_{{\rm copy},g}e_g
       \equiv\Ncopy\esite,\qquad
 \Ncopy=\dot N_{{\rm copy},0}+\dot N_{{\rm copy},g}.
 \label{eq:pinfo}
\end{equation}
For $\Ncopy>0$, $\esite$ is the rate-weighted net cost and can change with the process mixture; zero copying gives zero power.  Repair frequency enters the event count, whereas energy per event enters its cost.  Short-lived translation and ordinary protein renewal belong in $P_0$ or the production efficiency unless a separately specified heritable copying process is being counted.

%***********************************************
\section{Bound, dimensionless loads and copying closures}
\label{sec:bound}

Combining Eqs.~(\ref{eq:partition}), (\ref{eq:npp_def}) and (\ref{eq:pinfo}) gives the global material-productivity ceiling
\begin{equation}
  \NdotC \le \frac{\etabio}{\DGassim}
  \pos{\Pbio-\Pmaint-\Ncopy\esite},
  \label{eq:main_bound}
\end{equation}
with the corresponding areal form, where $\phimaint=\Pmaint/A_{\rm hab}$ and $\jcopy=\Ncopy/A_{\rm hab}$,
\begin{equation}
  \ndotC \le \frac{\etabio}{\DGassim}
  \pos{\phibio-\phimaint-\jcopy\esite}.
  \label{eq:areal_bound}
\end{equation}
The bound is conditional on the specified producer population and feasible mandatory costs.  Equality requires $P_u=0$.  The positive part is a convenient plotting convention: if $\Pmaint+\Pinf>\Pbio$ for a prescribed load, that population has no steady state even at zero NPP.  It is not made viable by setting its output to zero.

Two dimensionless loads are useful:
\begin{equation}
  \Lambda_{\rm info}=\frac{\Ncopy\esite}{\Pbio},
  \qquad
  \Theta_{\rm info}=\frac{\Ncopy\esite}{\Pbio-\Pmaint}.
  \label{eq:loads}
\end{equation}
$\Lambda_{\rm info}$ measures information processing relative to the total bioavailable budget.  For $P_{\rm res}>0$, $\Theta_{\rm info}$ measures information processing relative to the residual budget after environmental and basal overheads.  We call systems resource-dominated when $\Theta_{\rm info}\lesssim 0.1$, information-influenced when $0.1\lesssim\Theta_{\rm info}\lesssim0.5$, and information-limited when $0.5\lesssim\Theta_{\rm info}\le1$.  Prescribed loads with $\Theta_{\rm info}>1$ are infeasible.  The thresholds 0.1 and 0.5 are reporting conventions, not phase transitions.

The transition criterion is explicit:
\begin{equation}
  \Ncopy\esite \gtrsim \Theta_* (\Pbio-\Pmaint)
  = \Theta_* (1-f_{\rm maint})\Pbio,
  \label{eq:transition}
\end{equation}
where $\Theta_*\simeq0.1$ marks entry into the information-influenced regime and $\Theta_*\simeq0.5$ marks a large information allocation within the feasible region.  At fixed $f_{\rm maint}$ and $\Ncopy\esite$, the critical power is $\Pbio^{\rm crit}=\Ncopy\esite/[\Theta_*(1-f_{\rm maint})]$; at fixed $\Pmaint$ instead, it is $\Pmaint+\Ncopy\esite/\Theta_*$.

The copying closure determines whether an information burden persists as production declines.  For $B$ producer cells, genomic site inventory $L$ per cell and fractional repair-replacement rate $r$, a baseline model is $\dot N_{{\rm copy},0}=BLr$, accompanied by ordinary basal maintenance $P_0=Bp_0$.  Here $p_0$ is power per cell and $r$ has units s$^{-1}$.  Damage repair can persist at low growth \cite{Price2004,Hoehler2013}, but a fixed planetary burden requires $B$ and $r$ to remain approximately fixed on the timescale considered.  Supported biomass can instead decline with supply \cite{LaRowe2015}; that response is examined in Appendix~\ref{app:closure}.

For a microbial producer population whose biomass output is expressed in cell equivalents, let $m_C$ be carbon content in mol C cell$^{-1}$ and $\chi\ge1$ a specified multiplier for additional heritable copying per produced cell.  $L$ counts incorporated nucleotide sites, including both strands where appropriate, rather than ambiguously mixing sites and base pairs.  The growth-associated closure is
\begin{equation}
 \dot N_{{\rm copy},g}\simeq a\NdotC,
 \qquad a=\chi L/m_C.
 \label{eq:closure}
\end{equation}
Pure growth-coupled copying then gives
\begin{equation}
 \NdotC^{\rm max,cl}=
 \frac{\etabio P_{\rm res}}{\DGassim+\etabio a e_g},
 \qquad P_{\rm res}\ge0.
 \label{eq:closure_bound}
\end{equation}
The denominator correction is independent of absolute supply for fixed biological coefficients.  Low flux reduces NPP but does not by itself increase the copying fraction.

The mixed closure retains baseline and growth-coupled copying,
\begin{equation}
 \Ncopy=\dot N_{{\rm copy},0}+a\NdotC,
 \qquad \Pinf=\dot N_{{\rm copy},0}e_0+a\NdotC e_g,
 \label{eq:mixed_copy}
\end{equation}
and yields
\begin{equation}
 \NdotC^{\rm max,mix}=
 \frac{\etabio\pos{P_{\rm res}-\dot N_{{\rm copy},0}e_0}}
 {\DGassim+\etabio a e_g}.
 \label{eq:mixed_bound}
\end{equation}
This expression recovers the fixed-load model for negligible growth copying and Eq.~(\ref{eq:closure_bound}) for zero baseline.  The equality $e_0=e_g$ is not imposed: the two costs refer to different thermodynamic endpoints.

For $P_{\rm res}>0$, define $g=\DGassim/\etabio$ (J mol C$^{-1}$), $x_0=\dot N_{{\rm copy},0}e_0/P_{\rm res}$ and $q=\etabio a e_g/\DGassim$.  Relative to the zero-copying-dissipation ceiling $\NdotC^{(0)}=\etabio P_{\rm res}/\DGassim$, the feasible mixed model has
\begin{equation}
 \frac{\NdotC^{\rm max,mix}}{\NdotC^{(0)}}=\frac{1-x_0}{1+q},
 \qquad 0\le x_0\le1.
 \label{eq:relative_reduction}
\end{equation}
At capacity, $\Theta_{\rm info}=(x_0+q)/(1+q)$, exactly the fractional reduction relative to this \emph{fixed-population} control. A reduction of at least $\tau\in(0,1)$ occurs if and only if
\begin{equation}
 x_0+(1-\tau)q\ge\tau .
 \label{eq:threshold_exact}
\end{equation}
This inverse test specifies the required burden without selecting an unmeasured copying cost. For a feasible rectangle $x_-\le x_0\le x_+\le1$ and $0\le q_-\le q\le q_+$, monotonicity gives
\begin{equation}
 \frac{x_-+q_-}{1+q_-}\le\Theta_{\rm info}\le
 \frac{x_++q_+}{1+q_+}.
 \label{eq:interval_theta}
\end{equation}
These are exact bounds on the adopted domain, with no sampling weights. If input ranges cross $x_0=1$, feasibility must be reported separately. A population-adjusted comparison requires a different control, derived in Appendix~\ref{app:closure}.

%***********************************************
\section{Earth benchmark and planetary archetypes}
\label{sec:examples}

\subsection{Constrained resources and unresolved biological capture}

Table~\ref{tab:archetypes} collects resource quantities constrained at identifiable physical boundaries. Earth's global radiative budget gives absorbed solar power $1.225\times10^{17}$ W \cite{Stephens2012}. The Field et al.\ NPP synthesis, 104.9 Pg C yr$^{-1}$ \cite{Field1998}, corresponds to $1.30\times10^{14}$ W under the adopted $\DGassim=4.7\times10^5$ J mol C$^{-1}$ oxic reference \cite{Hoehler2023}. The ratio, $1.06\times10^{-3}$, constrains the aggregate conversion from that resource to retained biomass energy. It does not separately determine biological capture, basal maintenance, production efficiency or copying. Neither the NPP synthesis nor its chemical reference is assigned an invented observational error distribution.

\begin{table}[t]
\caption{Source-constrained planetary resources with distinct physical boundaries. None is a measurement of $\Pbio$. Intervals combine quoted marginal endpoints or geochemical scenarios and have no assigned joint confidence level. Appendix~\ref{app:parameters} gives the inputs and assumptions.}
\label{tab:archetypes}
\centering
\begin{tabular}{llll}
\toprule
Setting & Constrained resource & Power (W) & Unresolved conversion \\
\midrule
Earth & Absorbed broadband solar & $(1.214$--$1.236)\times10^{17}$ & Spectrum, delivery, capture \\
 &  radiation \cite{Stephens2012} &  & and biological allocations \\
\addlinespace

Temperate & Bolometric radiation intercepted by & $(0.891$--$1.015)\times10^{17}$ & Albedo, transmission, usable  \\
M-dwarf planet & TRAPPIST-1e \cite{Agol2021} &  & wavelengths and any biosphere \\
\addlinespace
Subsurface ocean & Enceladus escaping-H$_2$-equivalent & $(0.396$--$4.753)\times10^6$ & Reactant production, transport and  \\
&  methanogenesis \cite{Waite2017} &  & biological uptake \\
\addlinespace

Free-floating ocean  & Supplied redox fluxes $J_j$ and  & $\sum_jJ_j{\cal A}_j$ & No numerical flux inventory is  \\

planet & positive affinities ${\cal A}_j$ &  &  supplied \\
\bottomrule
\end{tabular}
\end{table}

For TRAPPIST-1e, measured-light-curve and stellar-model constraints determine intercepted radiation, not a surface ocean or its biological power \cite{Agol2021}. Ocean worlds instead require reactant fluxes and reaction affinities \cite{Hand2020,Hendrix2019}. Cassini-based H$_2$ escape and modeled methanogenesis affinities give a $0.396$--$4.753$ MW \emph{plume-equivalent} resource: the power released if an H$_2$ throughput of the escaping magnitude underwent that reaction. Escape is a residual flux; it neither measures biological consumption nor supplies an upper bound on total H$_2$ production. Quantitative ecosystem models likewise remain conditional on habitat and chemical partition \cite{Higgins2024}. For free-floating planets, thermal habitability \cite{Stevenson1999,Abbot2011,Badescu2011} does not determine a numerical redox supply. These examples retain the distinction between a constrained source and an assumed biological state.

If a resource $P_s$ is supplied to a specified habitat, write $Q=\Pbio-P_{\rm env}=\gamma P_s$, with $0\le\gamma\le1$. Let $M$ be total standing-population maintenance, including repair, and $m=M/Q$ for $Q>0$. At $Q=0$, any positive mandatory load is infeasible; the zero-load limit has zero production. With $K=a e_g$ and $q=K/g$, the feasible capacity satisfies
\begin{equation}
 \frac{g\NdotC^{\rm max}}{P_s}=\frac{\gamma(1-m)}{1+q},
 \qquad 0\le m\le1.
 \label{eq:source_propagation}
\end{equation}
This expression propagates each source interval through any independently justified biological ranges. With only physical fraction bounds, its lower bound is zero: planetary irradiation alone cannot supply a positive productivity prediction. The Enceladus row becomes a habitat supply only if the feed flux is explicitly equated to the plume-equivalent flux; it is not a bound on the entire ocean.

\subsection{A strain-matched inventory and inverse copying thresholds}
\label{sec:producer}

Genome length and cellular carbon cannot be sampled independently without checking material consistency. The sequenced primary producer \emph{Prochlorococcus} MED4 has $1{,}657{,}990$ base pairs \cite{Rocap2003}; measured cellular carbon is $45.8\pm4.0$ fg in P-replete cultures and $60.9\pm1.8$ fg in P-limited cultures \cite{Bertilsson2003}. For one chromosome equivalent, $L=3{,}315{,}980$ nucleotide sites and $\ell=L/m_C=(8.69,6.53)\times10^{20}$ sites mol C$^{-1}$ for the respective means. The culture conditions remain separate. Using their standing composition for newly produced cell equivalents is an assumption. Appendix~\ref{app:parameters} propagates the reported assay standard deviations and checks the DNA-carbon inventory.

Define $h_g=e_g\chi$ and normalize it by the published DNA fuel benchmark $w_D=9.5\times10^{-19}$ J site$^{-1}$ \cite{Kempes2017}, so $\lambda=h_g/w_D$ and $q=\etabio\ell w_D\lambda/\DGassim$. This normalization does not identify fuel expenditure with net dissipation. At $\lambda=1$, the physical restriction $\etabio\le1$ and the adopted oxic reference give growth-only ceiling reductions of at most 0.175\% and 0.132\% for the two mean inventories. Allowing the separate one-standard-deviation carbon intervals raises the larger value to 0.192\%. For a 10\% growth-only reduction, Eq.~(\ref{eq:threshold_exact}) instead requires
\begin{equation}
 h_g\ge\frac{\DGassim}{9\etabio\ell}.
 \label{eq:producer_threshold}
\end{equation}
At unit efficiency, the mean-inventory requirements are $\lambda\ge63.3$ and 84.1; at $\etabio=0.1$ they are ten times larger. These are reproducible conditional thresholds, not measured costs or a universal bound for other organisms. Additional chromosome copies, different biomass composition, repeated synthesis and different reference chemistry change the requirements explicitly.

\begin{figure}[t]
\centering
\includegraphics[width=\linewidth]{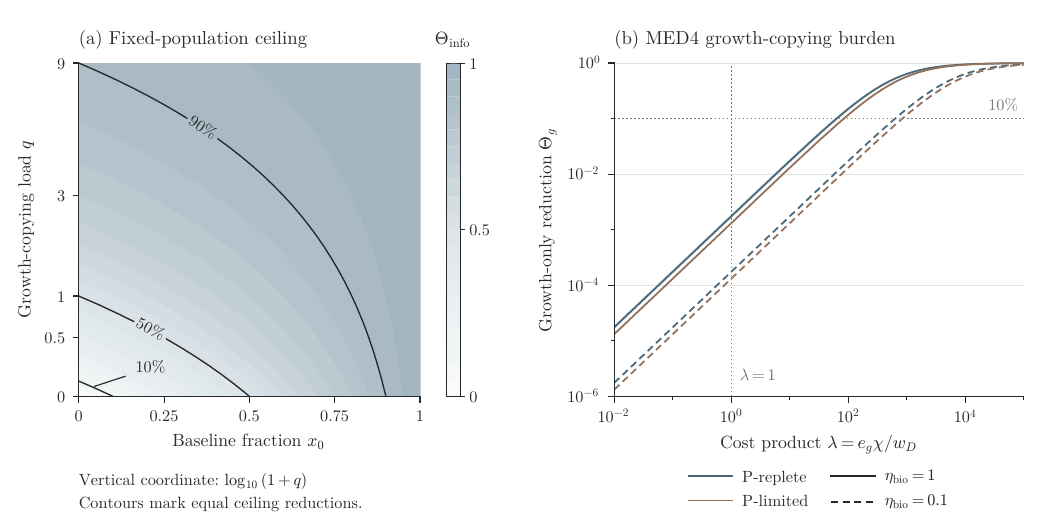}
\caption{Copying-induced reduction relative to the fixed-population zero-copying-dissipation ceiling. (a) $\Theta_{\rm info}=1-F$ over the feasible domain, with 10\%, 50\% and 90\% contours; the ordinate is $\log_{10}(1+q)$ with ticks labeled in $q$. (b) MED4 growth-only reduction $\Theta_g=q/(1+q)$ versus $\lambda=e_g\chi/w_D$. Colors distinguish culture conditions; solid and dashed curves use $\etabio=1$ and 0.1. Shaded bands propagate separate carbon means $\pm$ one sample SD at $\etabio=1$; they have no confidence-level interpretation. Dotted guides mark the fuel normalization $\lambda=1$ and the 10\% reduction threshold.}
\label{fig:regimes}
\end{figure}

\subsection{What maintenance measurements identify}
\label{sec:identifiability}

Maintenance and growth budgets depend on physiological state. In the chemolithoautotroph \emph{Methanococcus maripaludis}, lysis-corrected maintenance allocations range from 7\% at faster growth to 29\% at slower growth; these are flux-derived ATP-budget estimates, not isolated repair dissipation \cite{Mueller2021}. They cannot be transplanted as independent $P_0$ and copying priors.

If a consistently defined total maintenance power $M$ is constrained, write $P_0=(1-\varphi)M$ and $\dot N_{{\rm copy},0}e_0=\varphi M$, where $0\le\varphi\le1$ is the unknown repair share. Then
\begin{equation}
 \NdotC^{\rm max}=\frac{Q-M}{g+K},\qquad Q\ge M,
 \label{eq:total_maintenance}
\end{equation}
is exactly independent of $\varphi$. For $m<1$, the information share nevertheless spans
\begin{equation}
 \frac{q}{1+q}\le\Theta_{\rm info}\le\frac{m+q}{1+q}.
 \label{eq:allocation_interval}
\end{equation}
Thus total maintenance can constrain productivity without identifying the copying contribution. An empirical total growth cost similarly already includes $K$; adding $K$ again double counts it. Appendix~\ref{app:mc} combines exact parameter elasticities with global variance decompositions over declared domains. Near $x_0=1$ the fixed-population prediction is ill-conditioned; away from that boundary, threshold and interval statements remain directly testable.

%***********************************************

\section{Observational retrieval and consistency tests}
\label{sec:retrieval}

The observational role of the framework is a consistency test among productivity estimates, planetary power models and biological assumptions.  Geometry, stellar spectra, albedo and thermal emission constrain $\Pstar$ and $\Pabs$.  Climate, interior and geochemical models are required to infer $\PX$, $f_{\rm bio}$ and $P_{\rm env}$.  Producer biomass, basal maintenance, template turnover and the net costs $e_0,e_g$ remain biological inputs.  Their uncertainties and dependence on a shared population must be propagated together where such a model is available.

Atmospheric disequilibrium, temporal gas cycling, pigment-like surface features and coupled climate-geochemical models can inform productivity estimates, but do not directly measure NPP or copying dissipation \cite{KrissansenTotton2016,KrissansenTotton2018,Olson2018,Schwieterman2018,Kiang2007}.  For an assumed conversion efficiency, a proposed material production rate corresponds to
\begin{equation}
 \Pprod^{\rm obs}=\frac{\dot N_{{\rm C},{\rm obs}}\DGassim}{\etabio}.
 \label{eq:pobs}
\end{equation}
The fixed-load productive-power ceiling is
\begin{equation}
 \Pprod^{\rm max}=\pos{P_{\rm res}-\Ncopy\esite},
 \label{eq:pmax}
\end{equation}
with infeasible mandatory loads flagged separately.  For the mixed closure, the ceiling is $\DGassim\NdotC^{\rm max,mix}/\etabio$.  Equivalently, evaluate the full copying and maintenance demand at the proposed $\dot N_{{\rm C},{\rm obs}}$ and compare it with $\Pbio$; the two tests agree algebraically.

An apparent inconsistency identifies tension among the retrieval, energy-source model, overhead allocation and biological closure. It cannot isolate copying costs without independent constraints. A quantitative retrieval must propagate the same carbon measurement, conversion cost and biological population through both sides of the budget, using a joint distribution where observations support one. An arbitrary independent prior over genome size, biomass, repair and overhead is not such a distribution. We therefore report exact feasible intervals and design-space sensitivity, rather than numerical probabilities of planetary inconsistency. The useful observational outputs are a required source power, a maximum supportable load, or a threshold cost product whose empirical plausibility can be assessed.

%***********************************************

\section{Discussion and conclusions}
\label{sec:discussion}

The productivity ceiling is an energy-accounting constraint whose scientific content depends on a consistent producer boundary and a specified biological closure. The present development separates environmental and basal overheads, retained biomass free energy and net copying dissipation. It extends the ceiling in Ref.~\cite{Turyshev2026Pathway} through distinct maintenance and growth-copying costs, exact uncertainty bounds, and an explicit analysis of which cost allocations observations can identify.

Three conclusions are independent of assigned planetary biological powers. First, the normalized ceiling $(1-x_0)/(1+q)$ distinguishes a baseline burden from an added cost per unit production; Eq.~(\ref{eq:threshold_exact}) states exactly what combinations can produce a specified correction. Second, low energy flux alone does not increase a growth-associated fraction. A population that adjusts its biomass can also retain a constant repair fraction as supply falls, under fixed physiological coefficients. Third, changing the unknown repair share of an already constrained total maintenance budget cannot change absolute productivity, although it changes information attribution and the associated counterfactual.

The strain-matched producer calculation supplies a quantitative reference: the one-complement genome-to-carbon benchmark requires an effective copying cost $e_g\chi$ well above the fuel normalization for a 10\% growth-copying effect under the stated reference chemistry. It does not establish a universal biochemical cost. Likewise, observed radiative resources and plume chemistry constrain identifiable resource quantities without determining biological capture or biomass. The global sensitivity ranking changes with the declared domain and weighting; no universal dominant biological parameter follows from it. Exact elasticities instead show that residual power and mandatory maintenance require particular care near exhaustion.

The framework omits nutrient limitation, spatial ecology, episodic production and evolutionary changes in genome size or fidelity. It provides conditional upper envelopes, not realized NPP or evidence for an extraterrestrial biosphere. Its immediate value for comparative planetology is to reject internally inconsistent source--population combinations, quantify the biological requirements for a proposed productivity, and determine which measurements could distinguish copying dissipation from other energy demands. Net copying costs, producer-specific repair and independently constrained joint biological inputs remain necessary for calibrated planetary predictions.

%======= ACKNOWLEDGMENTS =================

\section*{Acknowledgments}
The work described here was carried out at the Jet Propulsion Laboratory, California Institute of Technology, Pasadena, California, under a contract with the National Aeronautics and Space Administration.  \textcopyright 2026. California Institute of Technology. Government sponsorship acknowledged.

\appendix
\suppressfloats[t]

\section{Parameter evidence and uncertainty propagation}
\label{app:parameters}

\subsection{Status of the adopted inputs}

Table~\ref{tab:parameters} separates measured quantities, observation-constrained estimates, physical restrictions and explicit modeling choices. A literature value is transferred only with its units, reference state and accounting boundary. The source package includes a parameter ledger, primary-source locations and full-precision arithmetic.

\begin{table}[t]
\caption{Input evidence and admissible treatment. An uncalibrated quantity has no adopted empirical distribution; it may still be constrained by a specified population and environment. The accompanying parameter ledger retains the full source details.}
\label{tab:parameters}
\centering
\begin{tabular}{llll}
\toprule
Quantity & Value or restriction & Evidence status & Use and limitation \\
\midrule
Planetary  & Table~\ref{tab:archetypes}; marginal endpoints  & Observation- & Resource envelopes do not identify   \\
 resources &  below & Observation-constrained; partly  &  biological capture \\
  &   &  model-derived &   \\
 
\addlinespace
$L$ for MED4 & $2(1{,}657{,}990)$ nucleotide sites \cite{Rocap2003} & Sequenced length; one & Both strands counted; extra copying is  \\
&  & chromosome equivalent  & distinct from retained chromosome  \\
&  & assumed & inventory \\

\addlinespace
Cellular C   & $45.8\pm4.0$ or $60.9\pm1.8$\ & Measured means and  & Separate culture conditions; triplicates  \\

for MED4 &  fg C cell$^{-1}$~\cite{Bertilsson2003} & sample SDs &  at  22$^\circ$C; no pooled biological prior \\

\addlinespace
$\DGassim$ & $4.7\times10^5$ J mol C$^{-1}$ & Adopted rounded oxic & Not measured for MED4 or universal in  \\
&  & reference \cite{Hoehler2023} &  anoxic environments \\

\addlinespace
$\etabio$, $g$ & $0<\etabio\le1$; $g=\DGassim/\etabio$ & Physical restriction; 0.1  & Empirical total growth yields already  \\
&  &  is a comparison & include copying costs \\
\addlinespace

$w_D$ & $9.5\times10^{-19}$ J site$^{-1}$ \cite{Kempes2017} & Published  & Normalizes $h_g=e_g\chi$; does not measure   \\
&  &  fuel-accounting scale &  net dissipation \\

\addlinespace
$e_g$, $\chi$ &  range & Physical signs and & Sweep the identifiable product $h_g$  \\
& range &  closure definition & without an empirical distribution \\

\addlinespace
$e_0$, $r$ & $e_0\ge\kb T\ln2\,\Isite$ for the specified  & Conditional  & Reconcile damage, repair length and  \\
 &  cycle; $r\ge0$ &  thermodynamic bound;  & dissipation; no universal repair prior \\
 &  &  repair unknown & \\

\addlinespace
$B$, $p_0$, $M$ & Nonnegative; mandatory costs  & Population-dependent  & Total maintenance includes repair;  \\
&  cannot exceed supply & constraints &  whole-microbial counts \cite{Whitman1998} are not  \\
&   &  &  producer-only biomass \\

\addlinespace
$\delta$ & Positive replacement rate in the  & Assumed closure  & Specified cell loss/export; molecular   \\

&  stationary closure &  coefficient & carbon  turnover is not a direct substitute \\

\addlinespace
$\gamma$, $m$, $\varphi$ &  budgets & Physical fraction  & No positive capture lower bound or  \\

&  budgets &  bounds & empirical distribution is assumed \\
\bottomrule
\end{tabular}
\end{table}

The rounded free-energy reference is consistent with the oxic carbon-oxidation scales used by Hoehler et al.\ \cite{Hoehler2023}, whose adopted values depend on biomass composition and process. It supplies a reproducible reference for the MED4 comparison, not an uncertainty distribution for $\DGassim$. Changing this reference rescales $q$ inversely and the required $h_g$ linearly. The anoxic ocean-world resource is therefore reported in W without silently applying that oxic carbon conversion.

The DNA fuel estimate in Ref.~\cite{Kempes2017} uses 12 ATP per nucleotide. Other accounts include precursor synthesis and opportunity costs \cite{Lynch2015}; their different boundaries do not define endpoints of one net-dissipation distribution. Likewise, very low whole-cell power estimates \cite{LaRowe2015,Hoehler2013} do not isolate producer-specific non-information maintenance. In \emph{M. maripaludis}, measured fluxes give growth-dependent maintenance and lysis corrections \cite{Mueller2021}. These observations support coupling the components of a budget, rather than varying separately charged totals independently.

\subsection{Planetary resource calculations}

For Earth, the global mean incoming and reflected solar fluxes are $340.2\pm0.1$ and $100.0\pm2$ W m$^{-2}$ in the observation-constrained 2000--2010 budget of Ref.~\cite{Stephens2012}. Using the explicitly adopted mean radius $6.371\times10^6$ m gives $P_{\rm abs,\star}=4\pi R_\oplus^2(340.2-100.0)=1.22517\times10^{17}$ W. Combining the quoted marginal endpoints yields $[1.21446,1.23589]\times10^{17}$ W. The factor of four is already included in global mean fluxes; these are multiplied by surface area, not planetary cross section. The envelope has no joint confidence interpretation. With 365.25 days per year and the stated rounded carbon conversion 12 g mol$^{-1}$, the NPP synthesis \cite{Field1998} gives $2.77007\times10^8$ mol C s$^{-1}$ and $1.30193\times10^{14}$ W for the adopted chemical reference.

For TRAPPIST-1e, Ref.~\cite{Agol2021} gives the absolute radius $R_p=(5.868^{+0.082}_{-0.075})\times10^6$ m and irradiance $S_\star=879\pm34$ W m$^{-2}$. Using the absolute quantities avoids mixing Earth-radius conventions. Equation~(\ref{eq:pstar}) gives $9.50865\times10^{16}$ W and the marginal-endpoint envelope $[8.90869\times10^{16},1.01544\times10^{17}]$ W. The radius and irradiance share inferred stellar parameters. A linearized propagation would obey
\begin{equation}
 \left(\frac{\sigma_P}{P}\right)^2\simeq
 4\left(\frac{\sigma_R}{R}\right)^2+
 \left(\frac{\sigma_S}{S}\right)^2+
 4\rho\frac{\sigma_R\sigma_S}{RS},
 \label{eq:source_covariance}
\end{equation}
where $\rho$ is their correlation. Using the mean of the two radius errors only for this linearized diagnostic gives 4.70\% if $\rho=0$, but 1.19--6.54\% over $-1\le\rho\le1$. A posterior interval requires the joint posterior; the endpoint calculation assumes no independence and claims no coverage probability.

For Enceladus, the inferred H$_2$ escape of $(1$--$5)\times10^9$ mol yr$^{-1}$ is conditional on a 200 kg s$^{-1}$ water-vapor plume \cite{Waite2017}. The modeled positive affinity ${\cal A}$ of
\begin{equation}
 \mathrm{CO_2+4H_2\longrightarrow CH_4+2H_2O}
\end{equation}
is about 50--120 kJ per mole of CH$_4$ reaction extent for the cited pH 9--11 scenarios at 273 K and 1 bar. Thus
\begin{equation}
 P_{\rm plume,eq}=\frac{F_{\rm H_2}}{4}{\cal A}
                 =0.396\text{--}4.753~\mathrm{MW}.
 \label{eq:enceladus_power}
\end{equation}
Here $F_{\rm H_2}$ is first converted to mol s$^{-1}$; division by four is required by reaction stoichiometry. Composition and affinity are related, so this rectangular scenario envelope is not a posterior distribution. At steady state, supply must support escaping gas plus any biotic and abiotic sinks. The escaping component does not identify their partition \cite{Waite2017,Higgins2024}.

For a free-floating habitat supplied by redox reactions, $P_{\rm chem}=\sum_jJ_j{\cal A}_j$ uses reaction flux $J_j$ in mol reaction s$^{-1}$ and affinity ${\cal A}_j$ in J mol reaction$^{-1}$. Shared reactants and coupled reactions must satisfy the same material balance; an energy source cannot be counted repeatedly. A thermal engine instead has a bound $P_{\rm int}(1-T_c/T_h)$ between specified hot and cold reservoirs. Neither expression determines biochemical access without transport and reaction constraints. No numerical internal power, ocean area or biological load is assigned to this case.

For positive source and conversion intervals, Eq.~(\ref{eq:source_propagation}) gives the enclosure
\begin{equation}
 \frac{P_{s,-}\gamma_-(1-m_+)}{g_+(1+q_+)}
 \le\NdotC^{\rm max}\le
 \frac{P_{s,+}\gamma_+(1-m_-)}{g_-(1+q_-)}.
 \label{eq:source_interval}
\end{equation}
The bounds are attained on an independent rectangular domain; for coupled inputs they remain conservative outer bounds when that domain contains the admissible set. They make explicit why a precise stellar resource does not compensate for an unconstrained capture fraction. Different planetary scenarios can be compared at specified common biological coefficients, but their resource ordering alone does not establish their productivity ordering.

\subsection{Strain-matched inventory and propagated assay variation}

The MED4 carbon assays are means and sample standard deviations of triplicates at 22$^\circ$C under constant illumination \cite{Bertilsson2003}. P-replete cells were sampled in mid-exponential growth, and P-limited cells in early stationary phase. The latter may include storage or cell-cycle effects. Using either standing inventory as the composition of newly produced cell equivalents is a scenario assumption, not a measured replication yield. A single chromosome complement is assumed; ploidy is not inferred from genome sequencing alone.

For carbon mass $c\pm s$ in fg per cell, the specified one-SD descriptive interval gives
\begin{equation}
 \ell=\frac{12L}{c\,10^{-15}},\qquad
 \ell_- =\frac{12L}{(c+s)10^{-15}},\quad
 \ell_+ =\frac{12L}{(c-s)10^{-15}}.
 \label{eq:assay_propagation}
\end{equation}
The first-order relative standard deviation is $\sigma_\ell/\ell\simeq s/c$, or 8.73\% and 2.96\% for the two assays. The endpoint ranges are conditional descriptive envelopes, not confidence intervals or known physical support. Table~\ref{tab:producer} reports the propagated quantities. No normal distribution is needed for these calculations.

\begin{table}[t]
\caption{MED4 reference calculations for the two separate culture conditions, assuming one dsDNA chromosome equivalent throughout. Bracketed values propagate each carbon mean $\pm$ one sample SD; they are descriptive ranges, not confidence intervals. The reduction $\Theta_g=q/(1+q)$ and the 10\% threshold $\lambda_{10\%}$ use $\etabio=1$ and $\DGassim=4.7\times10^5$ J mol C$^{-1}$. The cost product $\lambda=e_g\chi/w_D$ is normalized to a fuel benchmark, not a measured dissipation.}
\label{tab:producer}
\centering
\begin{tabular}{lrrcr}
\toprule
\toprule
Culture condition & ~~ \shortstack[r]{Cellular carbon\\(fg C cell$^{-1}$)} ~~& \shortstack[r]{$\ell$ ~~\\($10^{20}$ sites mol C$^{-1}$)} ~~& \multicolumn{1}{c}{\shortstack{$\Theta_g(\lambda=1)$\\(\%)}} & ~~\shortstack[r]{$\lambda_{10\%}$\\~~(dimensionless)} \\
\midrule
P-replete & $45.8\pm4.0$ ~~& $8.69$ ~~& 0.175 & $63.3$ \\
 & & $[7.99,\,9.52]$~~ & & $[57.7,\,68.8]$ \\
\addlinespace[4pt]
P-limited & $60.9\pm1.8$ ~~& $6.53$~~ & 0.132 & $84.1$ \\
 & & $[6.35,\,6.73]$~~ & & $[81.6,\,86.6]$ \\
\bottomrule
\end{tabular}
\end{table}

A necessary DNA-carbon inventory constraint is
\begin{equation}
 f_{{\rm DNA},C}=\frac{\bar c_{\rm nt}L}{N_A m_C}\le1,
 \qquad \bar c_{\rm nt}=10-\frac{f_{\rm GC}}2,
 \label{eq:inventory_constraint}
\end{equation}
where $N_A$ is Avogadro's constant and $f_{\rm GC}$ the GC fraction of complete dsDNA. An AT pair contains 20 carbon atoms and a GC pair 19, giving the stated mean per nucleotide. At MED4's sequenced GC fraction 0.308 \cite{Rocap2003}, the two mean inventories contain 1.42\% and 1.07\% of cellular carbon in DNA. Additional chromosome equivalents multiply this inventory; repeated synthesis followed by degradation belongs in $\chi$, not in retained DNA. This check rules out treating arbitrary genome-size and cell-carbon draws as automatically compatible.

For a required growth-only reduction $\tau$, the inverse result is
\begin{equation}
 \lambda_\tau=\frac{\tau\DGassim}
 {(1-\tau)\etabio\ell w_D}.
 \label{eq:lambda_threshold}
\end{equation}
At $\etabio=1$, the 50\% requirements are 569 and 757 for the mean inventories, with descriptive intervals [520,619] and [735,780]. All required $\lambda$ values scale as $\DGassim/\etabio$, and inversely with retained chromosome inventory. This explicit dependence is preferable to describing a selected $e_g$ or $\chi$ as a measured universal value.

\section{Systematic sensitivity and numerical validation}
\label{app:mc}

\subsection{Distribution-free sensitivity and feasibility}

Write $C_0=\dot N_{{\rm copy},0}e_0$, $R=P_{\rm res}$ and $K=a e_g$. In the feasible interior,
\begin{equation}
 Y\equiv\NdotC^{\rm max}=\frac{R-C_0}{g+K},\qquad
 x_0=C_0/R,\quad q=K/g.
 \label{eq:grouped_ceiling}
\end{equation}
The exact logarithmic elasticities, $E_z=\partial\ln Y/\partial\ln z$ with the other grouped inputs fixed, are
\begin{equation}
 E_R=\frac{1}{1-x_0},\qquad
 E_{C_0}=-\frac{x_0}{1-x_0},\qquad
 E_g=-\frac{1}{1+q},\qquad
 E_K=-\frac{q}{1+q}.
 \label{eq:elasticities}
\end{equation}
These formulas hold over the entire feasible interior, rather than only at a fiducial point. At $x_0=0.5,0.9,0.99$, the supply elasticity is 2, 10 and 100, respectively. No finite logarithmic bound exists as $x_0\to1$; finite envelopes require a specified distance from exhaustion. In contrast, the absolute derivatives
\begin{equation}
 \frac{\partial\Theta}{\partial x_0}=\frac{1}{1+q},\qquad
 \frac{\partial\Theta}{\partial q}=\frac{1-x_0}{(1+q)^2}
\end{equation}
are between zero and one. A nearly vanishing ceiling can therefore have a stable absolute allocation fraction but a highly uncertain relative NPP.

The same analysis can expose every biological dependency without independently sampling incompatible totals. Define carbon stock $M_C=Bm_C$, non-information basal cost per stock $s_0=p_0/m_C$, inventory $\ell=L/m_C$, maintenance cost-rate product $h_0=e_0r$, and growth cost product $h_g=e_g\chi$. Then
\begin{equation}
 D=Q-M_C(s_0+\ell h_0),\quad H=g+\ell h_g,\quad Y=D/H.
 \label{eq:primitive_model}
\end{equation}
Table~\ref{tab:elasticities} gives all seven elasticities. The shared $\ell$ couples baseline and growth costs, and $e_0,r$ or $e_g,\chi$ cannot be separately identified from their respective products in this budget. If only total maintenance per carbon stock is known, $m_s=s_0+\ell h_0$ replaces its components. Holding $m_s$ fixed makes the repair-allocation elasticity of absolute $Y$ exactly zero. Similarly, if a measured total growth cost $G_{\rm tot}=g+\ell h_g$ is used, the decomposition of that total cannot create an additional productivity penalty.

\begin{table}[t]
\caption{Exact log-elasticities for the grouped coordinates of Eq.~(\ref{eq:primitive_model}), at positive inputs and $D>0$; zero-cost entries use continuous limits. Each derivative holds the other coordinates fixed. A physiological inference must impose their actual covariances.}
\label{tab:elasticities}
\centering
\begin{tabular}{lll}
\toprule
Input $z$ & Meaning & $E_z=\partial\ln Y/\partial\ln z$ \\
\midrule
$Q$ & Supply after environmental overhead & $Q/D$ \\
$M_C$ & Standing producer carbon stock & $-M_C(s_0+\ell h_0)/D$ \\
$s_0$ & Basal power per unit carbon stock & $-M_Cs_0/D$ \\
$\ell$ & Template sites per unit carbon & $-M_C\ell h_0/D-\ell h_g/H$ \\
$h_0$ & Maintenance cost-rate product, $e_0r$ & $-M_C\ell h_0/D$ \\
$h_g$ & Growth cost product, $e_g\chi$ & $-\ell h_g/H$ \\
$g$ & Production cost excluding explicit copying & $-g/H$ \\
\bottomrule
\end{tabular}
\end{table}

The systematic response map uses 401 points in $x_0\in[0,1]$ and 401 points uniformly spaced in $\log_{10}(1+q)$ for $q\in[0,9]$. The upper $q$ endpoint represents a 90\% pure-growth correction; it is not a proposed physiological upper bound. Values beyond it follow the same exact asymptote. The MED4 sweep uses 600 logarithmic points in $\lambda=10^{-2}$--$10^5$ and 101 logarithmic points in $\etabio=10^{-4}$--1, together with the zero-cost and zero-efficiency limits. The interval $0<\etabio<10^{-4}$ is bounded by monotonicity, rather than assigned sampling weight. These ranges span the fuel-scale comparison and the 10\%, 50\% and 90\% requirements at the displayed efficiencies; they are diagnostic coordinates, not observations. Endpoint propagation is performed separately for the two measured carbon conditions.

\subsection{Global variance decomposition and its domain dependence}

For a declared independent design, Sobol indices decompose output variance into first-order terms and interactions \cite{Sobol2001}. We analyze the normalized ceiling $F=(1-x_0)/(1+q)$; $1-F=\Theta$ has identical indices. Set $U=1-x_0$, $V=(1+q)^{-1}$, $\mu_U=\mathbb E[U]$, $v_U=\operatorname{Var}(U)$ and similarly for $V$. Independence gives
\begin{align}
 \operatorname{Var}(F)&=v_U\mu_V^2+v_V\mu_U^2+v_Uv_V\equiv V_F,\\
 S_x&=\frac{v_U\mu_V^2}{V_F},&
 S_q&=\frac{v_V\mu_U^2}{V_F},\\
 T_x&=\frac{v_U\mathbb E[V^2]}{V_F},&
 T_q&=\frac{v_V\mathbb E[U^2]}{V_F}.
 \label{eq:sobol_exact}
\end{align}
The interaction index is $S_{xq}=v_Uv_V/V_F$. For $x_0$ uniform on $[0,a]$, $\mu_U=1-a/2$ and $v_U=a^2/12$. For $q$ uniform on $[0,b]$,
\begin{equation}
 \mathbb E[V]=\frac{\ln(1+b)}{b},\qquad
 \mathbb E[V^2]=\frac{1}{1+b}.
\end{equation}
For a design uniform in $z=\ln(1+q)$ on $[0,\ln(1+b)]$,
\begin{equation}
 \mathbb E[V]=\frac{1-(1+b)^{-1}}{\ln(1+b)},\qquad
 \mathbb E[V^2]=\frac{1-(1+b)^{-2}}{2\ln(1+b)}.
\end{equation}
These analytic moments determine every reported index; the measures are chosen design weights, not inferred biological probability densities.

\begin{table}[t]
\caption{Exact sensitivity indices for independent $x_0$ uniform on $[0,1]$ and the stated $q$ weightings. Endpoints $q_{\max}=1/9,1,9$ permit 10\%, 50\% and 90\% growth-only reductions. First-order indices $S_x,S_q$ exclude the interaction $S_{xq}$; total indices include it: $T_x=S_x+S_{xq}$ and $T_q=S_q+S_{xq}$. Domain and weighting choices preclude a universal biological ranking.}
\label{tab:sobol}
\centering
\begin{tabular}{llccccc}
\toprule
$q_{\max}$ ~~& ~~~Design weighting ~~~& \multicolumn{2}{c}{~~~First order~~~} & \multicolumn{2}{c}{~~~~Total~~~~} & \multicolumn{1}{c}{~~Interaction~~} \\
\cmidrule(lr){3-4}\cmidrule(lr){5-6}\cmidrule(l){7-7}
 & & \multicolumn{1}{c}{$S_x$} & \multicolumn{1}{c}{$S_q$} & \multicolumn{1}{c}{$T_x$} & \multicolumn{1}{c}{$T_q$} & \multicolumn{1}{c}{$S_{xq}$} \\
\midrule
$1/9$ & Uniform $q$ & 0.996 & 0.003 & 0.997 & 0.004 & 0.001 \\
$1/9$ & Uniform $\ln(1+q)$ & 0.996 & 0.003 & 0.997 & 0.004 & 0.001 \\
\addlinespace[3pt]
$1$ & Uniform $q$ & 0.860 & 0.105 & 0.895 & 0.140 & 0.035 \\
$1$ & Uniform $\ln(1+q)$ & 0.863 & 0.103 & 0.897 & 0.137 & 0.034 \\
\addlinespace[3pt]
$9$ & Uniform $q$ & 0.321 & 0.509 & 0.491 & 0.679 & 0.170 \\
$9$ & Uniform $\ln(1+q)$ & 0.380 & 0.465 & 0.535 & 0.620 & 0.155 \\
\bottomrule
\end{tabular}
\end{table}

The full baseline range and $q\le1/9$ yield $S_x=0.996$, whereas extending to $q\le9$ makes $S_q=0.509>S_x=0.321$ under uniform $q$ weighting. The reversal reflects the allowed effects. To control this asymmetry, the source package also evaluates matched domains $x_0\le\tau$, $q\le\tau/(1-\tau)$: at $\tau=0.1$, the uniform-design indices are 0.499 and 0.500; at $\tau=0.5$, 0.467 and 0.514. A further weighting check uses $z_g=q/(1+q)$, giving $\Theta=x_0+z_g-x_0z_g$. Identical independent uniform ranges for $x_0$ and $z_g$ give identical first-order and total indices by symmetry. These comparisons quantify how ranges and weights affect a variance-based ranking; the distribution-free thresholds do not change.

\begin{figure}[t]
\centering
\includegraphics[width=\linewidth]{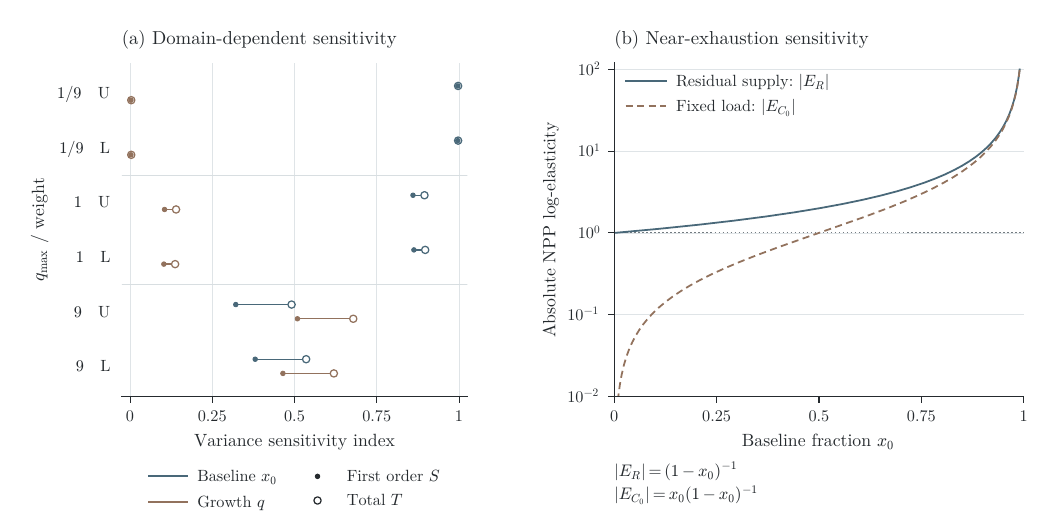}
\caption{Variance sensitivity and local conditioning. (a) Filled and open points give exact first-order $S$ and total $T$ indices for Table~\ref{tab:sobol}. Each connecting segment has length $T-S=S_{xq}$, the interaction contribution, not an uncertainty interval. U denotes uniform $q$ and L uniform $\ln(1+q)$; colors distinguish baseline and growth coordinates. (b) Magnitudes of the residual-supply and fixed-copying-load elasticities. $E_R>0$ and $E_{C_0}\le0$; both magnitudes diverge as $x_0\to1$. This diagnoses relative productivity uncertainty near budget exhaustion, not a physical singularity.}
\label{fig:mc}
\end{figure}

\subsection{Reproducibility and independent checks}

The supplied \texttt{calculations/uncertainty.py} generates the tables, figures, parameter sweeps and validation records without downloading data. A Jansen pick-freeze estimator \cite{Jansen1999,Saltelli2010} independently checks the analytic indices using two independent matrices of $2^{18}$ rows, NumPy's \texttt{PCG64(20261023)}, and four model evaluations per row. Prefixes of $2^{12}$, $2^{14}$ and $2^{16}$ rows document convergence. The maximum absolute index difference at $2^{18}$ is 0.00409; the tabulated indices themselves are analytic. Every sampled normalized state is feasible, so no rejection or conditioning introduces dependence between design coordinates. These simulations produce no probability of a real planet being information-limited.

Independent high-precision arithmetic reproduces the MED4 inverse thresholds, and separate source calculations reproduce the planetary resource conversions. The code tests $10^5$ feasible budgets spanning broad numerical scales; the largest relative energy-balance residual is $3.3\times10^{-16}$. It also verifies the two-cost reduction identity, both pure-closure limits, zero-cost and exhaustion limits, threshold equivalence, interval containment, the maintenance-partition invariance and the population-adjusted counterfactual. Central-difference derivatives reproduce the seven analytic elasticities to $1.0\times10^{-11}$ at the stated check point. Dimensional checks recover W in every power term and mol C s$^{-1}$ for productivity. The symmetric information channel gives zero mutual information at random-copy error and the stated correlation-loss value at perfect copying. A fixed load with $x_0>1$ is rejected as infeasible rather than reported as a viable zero-productivity population.

These checks validate the equations, implementations and specified numerical transformations. They do not independently validate extraterrestrial biomass, repair fluxes, net copying dissipation or biological capture. Those quantities remain unmeasured in the examples and are not generated by the sensitivity calculation.

\section{Control volume, feasibility and population adjustment}
\label{app:closure}

All terms use the producer boundary and one environmental free-energy reference. The retained flux $\NdotC\DGassim$ includes chemical and structural biomass energy; downstream consumer dissipation is excluded. Copying is charged only through excess dissipation. With $g=\DGassim/\etabio$, production's non-information dissipation is $(g-\DGassim)\NdotC$; thus $\etabio$ excludes the explicit basal and copying costs.

Let $B$ be the number of producer cells, $p_0$ their non-information basal power per cell, $L$ the number of genomic nucleotide sites copied per new cell, and $r$ the fractional genome-replacement rate due to repair.  Define $a=\chi L/m_C$, $b=e_0 Lr$, and $Q=\Pbio-P_{\rm env}$, where $e_0$ is the net dissipation per maintenance-cycle site and $e_g$ the excess dissipation per growth-associated site.  Here $m_C$ is in mol C cell$^{-1}$, $r$ is in s$^{-1}$, and $b$ and $p_0$ are in W cell$^{-1}$.  The budget, allowing unused supply $P_u\ge0$, is
\begin{equation}
 Q=B(p_0+b)+(g+e_g a)\NdotC+P_u .
 \label{eq:appendix_budget}
\end{equation}
This decomposition charges retained biomass, non-information dissipation and copying dissipation once each.  A correlation-loss lower bound may constrain $e_0$ for the specified complete maintenance cycle; it does not impose the same dissipation floor on $e_g$ for retained copies.

For fixed $B$, let $R=Q-p_0B>0$, $x_0=bB/R$, and $q=e_g a/g$.  If $0\le x_0\le1$, the ceiling and its information-load fraction obey
\begin{equation}
 \frac{\NdotC^{\rm max}}{\NdotC^{(0)}}
 =\frac{1-x_0}{1+q},\qquad
 \NdotC^{(0)}=\frac{R}{g},\qquad
 \Theta_{\rm info}^{\rm max}
 =\frac{x_0+q}{1+q}.
 \label{eq:appendix_reduction}
\end{equation}
The last identity follows by evaluating $(bB+e_g a\NdotC)/R$ at the ceiling.  It recovers the fixed-load limit $q=0$ and the growth-proportional limit $x_0=0$.  When $x_0>1$, the stipulated standing population is infeasible even at zero production; positive-part notation alone must not be interpreted as a viable steady state.

For a population-adjusted example, consider an idealized microbial producer population with constant loss rate $\delta>0$.  Assume all NPP produces new cellular biomass, with no organic-carbon imports, exudation or changing storage pool.  Stationarity of $m_CB$ then requires $\NdotC=\delta m_CB$.  Substitution into Eq.~(\ref{eq:appendix_budget}) at $P_u=0$ gives
\begin{equation}
 B_{\rm max}=\frac{Q}{p_0+e_0Lr+\delta m_Cg+e_g\chi L\delta},
 \qquad
 \NdotC^{\rm max}=\delta m_CB_{\rm max}.
 \label{eq:appendix_population}
\end{equation}
At this capacity state,
\begin{equation}
 \Theta_{\rm info}
 =\frac{e_0Lr+e_g\chi L\delta}
 {e_0Lr+e_g\chi L\delta+\delta m_Cg}.
 \label{eq:appendix_population_theta}
\end{equation}
For constant per-cell coefficients this ratio is independent of $Q$: reducing supply decreases population and productivity together. This allocation fraction differs from the loss relative to a separately re-equilibrated zero-copying-dissipation population. For that comparison, define $\ell=L/m_C$, $s_0=p_0/m_C$ and
\begin{equation}
 u=\frac{s_0}{\delta g},\qquad
 v=\frac{e_0\ell r}{\delta g},\qquad
 q=\frac{e_g\chi\ell}{g}.
\end{equation}
Then
\begin{equation}
 \dot N_{{\rm C},{\rm pop}}^{\rm max}=\frac{Q}{g(1+u+v+q)},\quad
 \frac{\dot N_{{\rm C},{\rm pop}}^{\rm max}}{\dot N_{{\rm C},{\rm pop}}^{(0)}}
 =\frac{1+u}{1+u+v+q},\quad
 \Theta_{\rm info}=\frac{v+q}{1+v+q},
 \label{eq:population_counterfactual}
\end{equation}
where $\dot N_{{\rm C},{\rm pop}}^{(0)}=Q/[g(1+u)]$. At fixed coefficients both fractions are independent of $Q$, but differ when $u>0$. Equation~(\ref{eq:relative_reduction}) applies only to fixed populations. Here $\delta$ describes cell death, export and washout, not necessarily carbon turnover. Physiology can change with supply.

%\bibliography{info-bio}

%apsrev4-2.bst 2019-01-14 (MD) hand-edited version of apsrev4-1.bst
%Control: key (0)
%Control: author (8) initials jnrlst
%Control: editor formatted (1) identically to author
%Control: production of article title (0) allowed
%Control: page (0) single
%Control: year (1) truncated
%Control: production of eprint (0) enabled
%

\end{document}